\documentclass[review,3p,10pt,times,authoryear]{elsarticle}

\usepackage{amssymb, amsmath}
\usepackage{bm, float}

\usepackage{hyperref}
\hypersetup{
    colorlinks  = true,     citecolor = blue
}

\newtheorem{pro}{Proposition}
\newproof{pf}{Proof}

\usepackage{graphicx} 
\graphicspath{{figures/}}
\usepackage{ulem}
\usepackage{enumitem}

\journal{A. Journal}

\begin{document}

\begin{frontmatter}

\title{Doubly-nonparametric generalized additive models}


\author{Alan Huang\corref{*}}
\ead{alan.huang@uq.edu.au}
\author{Nanxi Zhang\corref{}}
\cortext[*]{Corresponding author.}

\address{School of Mathematics and Physics, The University of Queensland, Qld 4072, Australia}

\begin{abstract}
The popular generalized additive model framework is extended to allow both the mean curves and the response distribution to be nonparametric. The approach is demonstrated to be a flexible yet parsimonious tool for data analysis in its own right, as well as being a useful tool for model selection and diagnosis in the classical generalized additive model framework. Finite-sample performance of the method is examined via various simulation settings and the method is illustrated on two data analysis examples.
\end{abstract}


\end{frontmatter}


\section{Introduction}
\label{sec:intro}
Generalized additive models \citep[GAMs,][]{HT1990} are popular nonparametric extensions of generalized linear models \citep[GLMs,][]{MN1989} that replace the linear predictor $\eta = \sum_{j=1}^d \beta_j X_j$ with a sum of smooth functional predictors $\eta = \sum_{j=1}^d f_j(X_j)$, where $\{f_j, j=1,2.\ldots, d\}$ is a set of smooth, but otherwise unspecified, functions and  $X_1, X_2, \ldots, X_d$ is a set of covariates. The flexibility of GAMs arise from their ability to model non-linear relationships between the response and the covariates without pre-specifying its form. The parsimony of GAMs comes from its additivity assumption, allowing each model component to be easily interpreted in a conditional manner, much like in classical linear regression and GLMs. For these reasons GAMs have found a wide range of applications in a variety of fields, including epidemiology \citep[e.g.,][]{S1994,HT1995} and ecology \citep[e.g.,][]{YM1991,GEH2002}.

As with GLMs, GAMs assume that the conditional responses come from some exponential family of distributions. This covers the popular normal, Poisson, binomial and gamma families, as well as many other less popular but equally useful families such as the inverse Gaussian for skewed continuous data, negative-binomial for over-dispersed counts, and the generalized Poisson \citep{F1993} and Conway-Maxwell-Poisson \citep{Huang2017} distributions for both over- and under-dispersed counts. Thus, the types of responses that can be covered by the GAM framework is very large. However, a particularly restrictive requirement of GAMs is that the response distribution needs to be correctly specified from the outset, with model misspecification typically leading to inefficient estimators and biased inferences on model parameters. It seems rather paradoxical to consider flexible curves for the mean function yet remain so rigid with the response distribution. Indeed, it is well-known that even in simple linear regression settings, misspecification of the response distribution can lead to significantly biased inferences \citep[e.g.,][]{E1967,W1982}. This problem is equally detrimental in nonparametric regression settings.

There are some existing methods that aim to relax the stringent distributional assumptions. For example, quasi-likelihood \citep[QL,][]{FHW1995} approaches require only a mean-variance relationship for the data. However, the first two moments still needs to be correctly specified, and this requirement is often too demanding in practice. An alternative approach is to model both the mean and variances nonparametrically \citep[e.g.,][Chapter 14.2]{RWC2003}, but this requires two levels of smoothing -- one level of smoothing for the mean function and another for the variance function. 
\citet{RS2005} extend this idea further by modelling the mean, scale, shape, skewness and kurtosis of the response conditional distribution. This is an incredibly flexible approach, but it requires multi-level smoothing which is somewhat unsuitable for smaller-sized problems. The proposed approach in this paper strikes a balance between the parsimony and interpretability of classical GAMs and the flexibility of nonparametric second (and higher) moment models of \citet{RWC2003} and \citet{RS2005}.

More precisely, this paper introduces a novel extension of classical GAMs that allows the response distribution to be unknown. That is, neither the form of the functional relationship between the response and the predictors, nor the distributional form of the response, need to correctly pre-specified. The proposed approach is a genuine extension of GLMs and GAMs, in that the only distributional assumption we make is that the data come from {\it some} exponential family -- but, crucially, we do not need to specify which exponential family a priori. The model space is in fact the class of {\it all} GAMs with a given set of additive predictors. 
%

An immediate advantage of the proposed approach is that we always remain in a full probability setting. In contrast, QL based methods typically do not correspond to actual probability models for the data and thus do not provide any further insight into the probabilistic mechanism generating the data beyond that of the first two moments. Having a full probability model is particularly useful for model selection and diagnosis, predictive inferences and nonparametric bootstrap resampling. Moreover, we also provide an explicit estimate of the underlying distribution, which we show in Section \ref{sec:asy} to be consistent and jointly asymptotically normal in distribution, along with the mean curves. Our approach can therefore also be used for model selection and diagnosis in the classical GAM framework. We illustrate how this can be carried out in the data analysis examples in Section \ref{sec:real}.

The only aspect of the proposed approach that requires user input is the selection of smoothing parameters. However, this process too can be automated via a selection method such as cross-validation. Note that smoothing parameters are central to all smoothing methods in statistics, including the classical approach of \citet{HT1990} which this paper extends. An attractive aspect of our proposed approach is that it only requires specification of the smoothing parameters, whereas existing methods require specification of both the smoothing parameters and an underlying response distribution or variance and higher order moments for the data. As is evidenced through the various simulation examples in Section \ref{sec:sims} and the data analysis examples in Section \ref{sec:real}, relaxing the distributional assumptions in GAMs makes the doubly-nonparametric GAM approach a particularly flexible yet parsimonious tool for regression analyses.

\section{Model and method}
\label{sec:model}
\subsection{Classical nonparametric GAMs}
\label{sec:cpgam}
We first review the classical penalized likelihood approach to nonparametric GAMs. The extension to doubly-nonparametric GAMs is then developed using a novel exponential tilt representation of GLMs introduced in \citet{RG2009}. 

Following \citet[][Chapter 6]{HT1990}, recall that a GAM assumes that the conditional mean $\mu_i = E(Y_i|\boldsymbol{X}_i)$ of a response $Y_i$ is related to a corresponding set of predictors $\boldsymbol{X}_i = (X_{i1}, X_{i2}, \ldots, X_{id})^T$ via $g(\mu_i) = \eta_i = \sum_{j=1}^d f_j(X_{ij})$, where $g(\cdot)$ is a user-specified link function as in classical GLMs. In a slight abuse of notation, it is often convenient to write $\mu(\cdot) = g^{-1}(\cdot)$ for the inverse-link function, so that $\mu_i = \mu\big(\sum_{j=1}^d f_j(X_{ij})\big)$. It is also assumed that the conditional distribution of each response $Y_i$, given the predictors $\boldsymbol{X}_i$, comes from some exponential family of distributions with densities (with respect to some dominating measure) of the form
\begin{equation}
dF_i(y) = \exp\left\{ \frac{y\theta_i - b(\theta_i)}{\phi} + c(y; \phi)\right\} \ ,
\label{eq:expfam}
\end{equation}
where $\phi$ is a scale parameter, and $b(\cdot)$ and $c(\cdot)$ are known functions that determine the form of the distributions. Note that the means $\mu_i = \int y dF_i(y) = b'(\theta_i)$ are related to the canonical parameter $\theta_i$ via the canonical link $b'(\cdot)$. In turn, each $\theta_i$ is related to the predictors $\boldsymbol{X}_i$ via $g(b'(\theta_i)) = \sum_{j=1}^d f_j(X_{ij})$. If the canonical link is used, then $g^{-1} = b'$ and so $\theta_i = \sum_{j=1}^d f_j(X_{ij})$ directly. However, $g$ needs not be the canonical link in general. For example, for count responses it is ordinarily sensible to use the log-link, $g(\cdot) = \log(\cdot)$, regardless of whether the underlying distribution in (\ref{eq:expfam}) is Poisson, negative-binomial, generalized Poisson or Conway-Maxwell-Poisson. This allows each function $f_j$ to be interpreted directly in terms of a multiplicative effect on the mean response, irrespective of the underlying distribution.

Various computational approaches are available for fitting each individual function and the overall mean curve to data, including but not limited to back-fitting combined with local scoring \citep{HT1990}, marginal integration approach \citep{LN1995} and low-rank smoothers \citep{ME1998}. Here we focus on the penalized splines technique in which each smooth function $f_j$ can be approximated using regression splines, that is, 
$$
f_j(\cdot) = \sum_{k = 1}^K \beta_{jk} B_k(\cdot)\ ,
$$
where $\boldsymbol{B} = \left(B_1, B_2, \ldots, B_K\right)^T$ is a set of basis functions, such as B-splines or P-splines \citep[see, e.g.,][]{RWC2003, W2017}, and $\bm{\beta}_j = (\beta_{j1}, \beta_{j2},\ldots, \beta_{jK})^T$ is a corresponding vector of coefficients. For convenience, write $\bm{\beta} = (\bm{\beta}_1^T,\ldots,\bm{\beta}_d^T)^T$ for the full vector of coefficients and extend $\bm{B}$ to be the multivariate function $\bm{B}(\bm{X}_i) = (\bm{B}^T(X_{i1}),\bm{B}^T(X_{i2}),\ldots,\bm{B}^T(X_{id}))^T$ so that each functional predictor can be written as $\eta_i = \sum_{j=1}^d f_j(X_{ij}) = \bm{\beta}^T \bm{B}(\bm{X}_i)$.

Given a set of observations $(\boldsymbol{X}_1, Y_1), (\boldsymbol{X}_2, Y_2), \ldots, (\boldsymbol{X}_n, Y_n)$, the penalized maximum likelihood estimator of $\bm{\beta}$ can then be obtained by maximizing the penalized log-likelihood function,
$$
\ell_{n\lambda}(\bm{\beta}) = \ell_n(\bm{\beta}) - \frac{1}{2}\sum_{j = 1}^d \lambda_j \bm{\beta}_j^T D \bm{\beta}_j \ ,
$$
where $\ell_n(\bm{\beta}) = \frac{1}{n}\sum_{i=1}^n [Y_i \theta_i - b(\theta_i)]$ is the unscaled log-likelihood with $\theta_i \equiv \theta_i(\bm{\beta})$ given by $g(b'(\theta_i)) = \bm{\beta}^T \bm{B}(\bm{X}_i)$, $\lambda_j \geq 0$ are smoothing parameters, 
and $D$ is some $K \times K$ positive semi-definite symmetric penalty matrix.
Writing $P = \mbox{diag}(\lambda_1 D, \lambda_2 D, \ldots, \lambda_d D)$ for the block-diagonal matrix with $\lambda_j D$ on the diagonals, the penalized log-likelihood can then be written as
\begin{equation}
\ell_{n\lambda}(\bm{\beta}) = \ell_n(\bm{\beta}) - \frac{1}{2}\bm{\beta}^T P \bm{\beta}\ .
\label{eq:pgam}
\end{equation}

The penalty term in (\ref{eq:pgam}) controls the roughness of each function to avoid over-fitting. A smaller value of $\lambda_j$ results in a more wiggly fitted function $\hat{f}_j$ that may capture local fluctuations, whereas increasing the value of $\lambda_j$ leads to an increasingly linear estimation of function $f_j$. How this translates to the smoothness of the mean curve depends on the user-specified link function, $g$. For model identifiability, each smooth function $f_j$ can be constrained to sum to 0, i.e., $\sum_{i=1}^n f_j(X_{ij}) = 0$ for $j=1,\ldots, d$. For more discussions on the theoretical and practical properties of penalized regression splines, see \citet{HT1990} and \citet{W2017}.

\subsection{Doubly-nonparametric GAMs}
\label{sec:spgam}
A recent innovation by \citet{RG2009} showed that any family of distributions with densities of the form (\ref{eq:expfam}) can be rewritten as $dF_i(y) = \exp{\{\theta_i y  - b_i \}} dF(y)$ for some reference distribution $F$, where the cumulant generating function $b_i \equiv b(X_i; \bm{\beta}, F)$ and canonical parameter $\theta_i \equiv \theta(X_i; \bm{\beta}, F)$ are given by the joint solution to the normalization constraint,
\begin{equation}
\int_\mathcal{Y} \exp\{\theta(X_i; \bm{\beta}, F) y - b(X_i; \bm{\beta}, F)\} dF(y) \ = \ 1 \ ,
\label{eq:norm}
\end{equation}
and the mean constraint,
\begin{equation}
\int_\mathcal{Y} y\exp\{\theta(X_i; \bm{\beta}, F) y - b(X_i; \bm{\beta}, F)\} dF(y) \ = \  \mu\big(\bm{\beta}^T \bm{B}(\bm{X}_i) \big) \ .
\label{eq:mean}
\end{equation}
In other words, each density $dF_i$ is an exponential tilt of some reference density $dF$, with the amount of tilting $\theta_i$ determined by the mean $\mu_i = \mu\big(\bm{\beta}^T \bm{B}(\bm{X}_i) \big)$. Note that the scale parameter $\phi$ has been absorbed into the functions $b, \theta$ and $F$. 

The key advantage of the exponential tilt representation (\ref{eq:norm})--(\ref{eq:mean}) is that it allows the underlying response distribution $F$ itself to be considered as a parameter in the model. Indeed, we can now write the penalized log-likelihood (\ref{eq:pgam}) as a function of both $\bm{\beta}$ and $F$,
\begin{equation}
\ell_{n \lambda}(\bm{\beta}, F) =  \frac{1}{n} \sum_{i=1}^n \{ \log dF(Y_i) - b(X_i;\bm{\beta},F) + \theta(X_i; \bm{\beta}, F) Y_i \} - \frac{1}{2}\bm{\beta}^T P \bm{\beta} \ .
\label{eq:dnpgam}
\end{equation}

The GAM characterized by the log-likelihood (\ref{eq:dnpgam}) is now doubly-nonparametric, as the parameter space for $F$ is the infinite-dimensional space of all distributions having a Laplace transform in some neighbourhood of 0. This covers all discrete and continuous exponential families, including the Poisson, Generalized Poisson and Conway-Maxwell-Poisson families for discrete data and the normal, gamma and inverse-Gaussian families for continuous data.
Note that the requirement of a Laplace transform is needed so that the cumulant generating function $b(\cdot)$ in (\ref{eq:norm}) is well-defined. 

Treating $F$ as a free parameter introduces much flexibility and robustness into the model. For example, over-dispersed counts can be dealt with simply by $F$ having heavier tails than a Poisson distribution, while under-dispersed counts can be dealt with simply by $F$ having lighter tails than a Poisson distribution. Similarly, zero-inflated counts can be dealt with simply by $F$ having excess probability mass at zero. More importantly, perhaps, is that $F$ can be left completely unspecified and estimated nonparametrically from the data along with the mean curves. In other words, we can let the data inform us which mean curves and response distribution fit best.

The seemingly intractable problem of working with this infinite-dimensional distributional space can be reduced to a finite maximization problem via constructing an empirical likelihood by replacing the density $dF$ with a set of non-negative probability masses $\bm{p} = (p_1, p_2,\ldots, p_n)^T$, so that  $F(y) = \sum_{i=1}^n p_i I(Y_i \le y)$, where $I$ is the indicator function. A doubly-nonparametric penalized maximum likelihood estimator for $\bm{\beta}$ and $\bm{p}$ can then be defined as the solution to the finite constrained optimization problem:

\noindent
\hrulefill
\texttt{
\begin{eqnarray*}
&\mbox{maximize }& \ell_{n\lambda} = \frac{1}{n} \sum_{i=1}^n \left\{ \log(p_i) - b_i + \theta_i Y_i \right\} - \frac{1}{2}\bm{\beta}^T P \bm{\beta} \quad \mbox{ in } \ \bm{\beta}, \bm{p}, \bm{b} \mbox{ and } \bm{\theta} \ , \\
&\mbox{subject to }& \sum_{j=1}^n \exp\{\theta_i Y_j - b_i \} p_j = 1 \ , \mbox{ for } i=1,2,\ldots,n \ , \\
&\mbox{ and } & \sum_{j=1}^n Y_j \exp\{\theta_i Y_j- b_i \} p_j = \mu\big(\bm{\beta}^T \bm{B}(\bm{X}_i) \big)\ , \mbox{ for } i=1,2,\ldots, n \ .
\end{eqnarray*}
}
\hrulefill

Denoting the maximizer by $\hat{\bm{\beta}}$ and $\hat{\bm{p}}$, the penalized maximum likelihood estimator of the underlying distribution $F$ is then given by $\hat F(y) = \sum_{i=1}^n \hat p_i I(Y_i \le y)$. In the next section, we show that $(\hat{\bm{\beta}}, \hat F)$ is consistent and jointly asymptotically normal in distribution. This allows us to construct asymptotically correct confidence bands for each function predictor $f_j$ and to develop model diagnostics for the distributional component $F$.

The exponential tilt representation (\ref{eq:norm})--(\ref{eq:mean}) is also used in \citet{Huang2014} to develop a semiparametric extension of GLMs in which the mean function is parametric but the response distribution is nonparametric. The key innovation in this paper is that both the mean function and the response distribution can be modelled nonparametrically, allowing the data to ``speak for themselves" in a doubly-nonparametric way. The close connection between doubly-nonparametric GAMs and semiparametric GLMs makes the corresponding techniques and arguments in \citet{Huang2014} readily applicable for our proposed method. We use these methods in deriving the asymptotic properties of doubly-nonparametric estimator in Section \ref{sec:asy}.

\section{Asymptotic theory}
\label{sec:asy}
For the joint parameter space, define a distance function by $\|(\bm{\beta}_1, F_1) - (\bm{\beta}_2, F_2)\| = \|\bm{\beta}_1 - \bm{\beta}_2\| + \|F_1 - F_2\|_{\mathcal{H}_L}$, where $\|\bm{\beta}_1 - \bm{\beta}_2\|$ is the Euclidean distance and $\|F_1 - F_2\|_{\mathcal{H}_L} = \sup_{h \in \mathcal{H}_L} \int h(dF_1 - dF_2)$ with $\mathcal{H}_L:=\{I(y \leq r): r \in \mathcal{Y}\}$ is the set of all left indicator functions on $\mathcal{Y}$. We will use this distance function when establishing the asymptotic properties of the proposed doubly-nonparametirc GAM estimator.

Let $(\bm{X}, Y)$ be a generic observation pair. Following the derivations in \citet{Huang2014}, the penalized score function for $\bm{\beta}$ has the form
\begin{eqnarray*}
S^\lambda_{\bm{\beta}, F}(\bm{X},Y) 
&=& \left[Y-\mu(\bm{\beta}^T B(\bm{X})) \right] \frac{\mu'(\bm{\beta}^T B(\bm{X}))}{V(\bm{X}; \bm{\beta}, F)}B(\bm{X}) - P \bm{\beta} \ ,
\end{eqnarray*}
where 
$
V(\bm{X}; \bm{\beta}, F) = \int_\mathcal{Y} \big[Y-\mu(\bm{\beta}^T B(\bm{X})) \big]^2 \exp\left\{ \theta(\bm{X};\bm{\beta}, F)y - b(\bm{X};\bm{\beta}, F) \right\} dF(y)
$
is the conditional variance of $Y$ given $X$. Similarly, a score operator $A_{\bm{\beta}, F}: \mathcal{H}_L \to l^\infty(\mathcal{H}_L)$ for the distribution parameter $F$ can be derived as
\begin{eqnarray*}
A_{\bm{\beta}, F}h(X,Y) = h(Y) - B_{\bm{\beta}, F}h(X) - \frac{Y-\mu(\bm{\beta}^T B(\bm{X}))}{\sqrt{V(X; \bm{\beta}, F)}}C_{\bm{\beta}, F}h(X) \ ,
\end{eqnarray*}
where
$
B_{\bm{\beta}, F}h(X) = E_{\bm{\beta}, F} \left[h(Y)|X \right] 
$
and 
$
C_{\bm{\beta}, F}h(X) = E_{\bm{\beta}, F}\big[h(Y)(Y-\mu(\bm{\beta}^T B(\bm{X})))|X \big] \big/\sqrt{V(X; \bm{\beta}, F)}
$; see \citet[][Section 3]{Huang2014} for more details of these calculations.

The doubly-nonparametric maximum penalized likelihood estimator $(\hat{\bm{\beta}}, \hat F)$ can then be characterized as the joint solution to the score equations
$$
\frac{1}{n}\sum_{i=1}^n S_{\bm{\beta}, F}^\lambda (\bm{X}_i, Y_i) =0 \quad \mbox{ and } \quad \frac{1}{n}\sum_{i=1}^n A_{\bm{\beta}, F} h (\bm{X}_i, Y_i) = 0 \ .
$$
This characterization proves useful for establishing the consistency and joint asymptotic normality of the proposed estimator. The proof of Proposition 1 below is given in the Online Supplement.


%

\begin{pro}[Consistency and joint asymptotic normality]
Under Assumptions A1--A3 in the Appendix, 
\begin{enumerate}
\item[(a)] if the smoothing parameters satisfy $\lambda_j = o(1)$, then there exists a local maximizer $(\hat{\bm{\beta}}, \hat{F})$ of (\ref{eq:dnpgam}) such that $\hat{\bm{\beta}} \rightarrow \bm{\beta}$ in probability and $\hat{F} \rightarrow F$ in probability relative to the weak topology; 
\item[(b)] if the smoothing parameters satisfy $\lambda_j = o(n^{-1/2})$, then
\begin{equation*}
\sqrt{n} 
\begin{pmatrix}
\hat{\bm{\beta}} - \bm{\beta} \\
\hat{F} - F
\end{pmatrix}
\rightarrow
\begin{pmatrix}
G_{\bm{\beta}} \\
G_F
\end{pmatrix}
\end{equation*}
in distribution in $\mathbb{R}^{Kd} \times l^\infty(\mathcal{H}_L)$, where $G_{\bm{\beta}}$ is a mean zero normal random vector with covariance matrix
$$
W_{\bm{\beta}} = \left\{ E_X \left( \frac{\mu'(\bm{\beta}^T B(\bm{X}))^2 B(X)^T B(X)}{V(X; \bm{\beta}, F)} \right) \right\}^{-1},
$$
$G_F$ is a mean zero Gaussian random process indexed by $h \in \mathcal{H}_L$ with some covariance function $W_F(h_1,h_2)$ given in the Online Supplement, and $G_{\bm{\beta}}$ and $G_F$ are independent. 
\end{enumerate}
\end{pro}

The asymptotic independence of $G_{\bm{\beta}}$ and $G_F$ motivates a simple estimator of the covariance matrix of $\hat{\bm{\beta}}$ for given $\hat F$ using a sandwich formula. Let
$$
{W}(\bm{\beta}) = \sum_{i=1}^n \frac{\partial}{\partial \bm{\beta}^T} S^\lambda_{\bm{\beta}, \hat F}(\bm{X}_i,Y_i) \quad \mbox{and} \quad {H}(\bm{\beta}) = \sum_{i=1}^n S^\lambda_{\bm{\beta}, \hat F}(\bm{X}_i,Y_i) S^\lambda_{\bm{\beta}, \hat F}(\bm{X}_i,Y_i)^T \ .
$$
Then the covariance matrix of $\hat{\bm{\beta}}$ can be estimated by 
\begin{equation}
\hat W_{\bm{\beta}} = {W}(\hat{\bm{\beta}})^{-1} {H}(\hat{\bm{\beta}}) {W}(\hat{\bm{\beta}})^{-T} \ . 
\label{eq:sandwich}
\end{equation}
This empirical covariance matrix can be used for inferences on each smooth function and the overall mean curve. More specifically, writing the estimated smooth predictors as $\hat{f}_j = \hat{\bm{\beta}}_j^T \bm{B}$, an approximate 95\% confidence band for each smooth function can be obtained by $\hat f_j \pm 1.96 se(\hat f_j)$, where $se(\hat f_j) = \sqrt{\bm{B}^T \hat{W}_{\bm{\beta}_j} \bm{B}}$ and $\hat{W}_{\bm{\beta}_j}$ is the $j$-th $K \times K$ block matrix along the diagonal of matrix $\hat{W}_{\bm{\beta}}$. Similarly, the additive predictor can be estimated by $\hat \eta = \sum_{j=1}^d \hat f_j$ with estimated standard error $se(\hat \eta) = \sqrt{\bm{B}^T \hat W_\beta \bm{B}}$, so that an approximate 95\% confidence band for the overall mean curve can be constructed via $g^{-1}(\hat \eta \pm 1.96 se(\hat \eta))$ .

To assess the goodness-of-fit of the estimated distribution $\hat F$, we recommend using a probability inverse transform \citep[PIT,][]{S1985}. If the fitted model is indeed appropriate, then the PIT should resemble a random sample from a standard uniform distribution. This can be assessed graphically using either a histogram or a quantile-quantile plot of the PIT against the uniform distribution. The estimated distribution $\hat F$ can also be directly plotted, perhaps alongside a postulated parametric model. These plots can then be used for model selection and diagnosis in the classical GAM setting. Some examples of these plots can be found in Sections \ref{sec:sims} and \ref{sec:real}.


Throughout the rest of this paper we treat the smoothing parameters $\lambda_j$ as being given sequences. In practice, there are a few competing ways to choose the smoothing parameter, with perhaps the most popular approach being cross-validation. Our recommendation for the doubly-nonparametric GAM framework is to simply plug in the default smoothing parameters obtained from fitting a preliminary \texttt{gam} from the \texttt{mgcv} R package \citep{W2016} under some working distribution model using generalized cross-validation. Although the smoothing parameters chosen in this way might be different to the ``optimal" set of smoothing parameters for any given problem, we find that this simple plug-in approach still enjoys excellent performance in all our simulations and data analysis examples. In fact, the smoothing parameters turn out to be rather robust to the working distributional model. This is an advantage of carrying out smoothing on the mean scale rather than on the canonical scale, as the latter depends critically on the underlying distribution.

\section{Simulation study}
\label{sec:sims}
The doubly-nonparametric GAM (\ref{eq:norm})--(\ref{eq:dnpgam}) is an extension of classical GAMs that does not require correct specification of the conditional distribution or variance function for the response. Thus, the approach is expected to be flexible enough to handle a very wide range of response types. Here, we examine the practical performance of the proposed approach using various simulations. We adopt the design from \citet{MW2012}. Consider the following set of smooth functions,
\begin{equation}
f_1(x_1) = 2 \sin (\pi x_1) \ , \quad f_2(x_2) = e^{2x_2} \ , \quad f_3(x_3) = x_3^{11}\{10(1-x_3)\}^6 + 10(10x_3)^3(1-x_3)^{10} \ , \quad f_4(x_4) = 0 \ .
\label{eq:f}
\end{equation}
The corresponding covariates $X_1,\ldots,X_4$ are each generated independently from $U(0,1)$. The additive predictor $\eta = f_1(X_1) + f_2(X_2) + f_3(X_3) + f_4(X_4)$ is then transformed via the inverse-link to generate the true mean curve $\mu = g^{-1}(\eta)$.


To examine the flexibility, robustness and practical performance of the proposed approach in both correctly specified and misspecified scenarios, we simulated data from a range of distributional settings. These include the normal and gamma distributions for continuous data, the Poisson, over-dispersed negative-binomial, and both over-dispersed and under-dispersed mean-parametrized Conway-Maxwell-Poisson distributions \citep{Huang2017} for discrete data, and the binomial and quasi-binomial distributions for binary data.
Table \ref{tab:setting} summarizes the simulation settings considered in this paper.

\begin{table}[]
\centering
\footnotesize
\caption{Simulation settings -- conditional distributions, mean functions and variance functions for $Y|\bm{X}$. In all scenarios, the additive predictor is $\eta = f_1(X_1) + f_2(X_2) + f_3(X_3) + f_4(X_4)$ with the smooth functions $f_1, f_2, f_3$ and $f_4$ specified in (\ref{eq:f}). \newline}
\begin{tabular}{llccl}
\hline
response type & conditional distribution & mean $\mu$ & dispersion & variance \\
\hline
continuous & 1. Gamma & $\exp(\eta)$ & 0.6 & $0.6 \mu^2$ \\
& 2. Heteroscedastic Normal & $\exp(\eta)$ & -- & $\mu$ \\
count & 3. Poisson & $\exp(\eta)$ & -- & $\mu$ \\
& 4. Negative-Binomial (over-dispersed) & $\exp(\eta)$ & 1  & $\mu + \mu^2$ \\
& 5. Conway-Maxwell-Poisson (under-dispersed)& $\exp(\eta)$ & 3 & no closed form \\
& 6. Conway-Maxwell-Poisson (over-dispersed)& $\exp(\eta)$ & 0.2 & no closed form \\
binary & 7. Binomial (with 3 trials) & $ \frac{\exp(\eta)}{1+\exp(\eta)}$ & -- & $\mu(1-\mu)$ \\
& 8. Quasi-Binomial (with 6 trials) & $\frac{\exp(\eta)}{1+\exp(\eta)}$ & 4 & $4\mu(1-\mu)$ \\
\hline
\end{tabular}
\label{tab:setting}
\end{table}


Settings 1, 3, 4 and 7 correspond to ``standard" GAM scenarios, and the correct model can be fit using existing software such as the $\texttt{gam}$ function in the $\texttt{mgcv}$ R package \citep{W2016}. In contrast, settings 2, 5, 6 and 8 are non-standard models, and the correct model cannot be easily fit using any existing software. Although setting 2 has the form of a quasi-Poisson model with the log-link, the observations themselves can be negative and software such as $\texttt{gam}$ cannot fit quasi-Poisson models when there are negative observations. Moreover, setting 2 cannot be written in the exponential family form (\ref{eq:expfam}), so it is outside our model space and is therefore misspecified. Note that Settings 4, 5 and 6 are also not of the exponential family form (\ref{eq:expfam}) unless the dispersion parameter is known {\it a priori} -- these settings can also be considered as being misspecified. It is precisely these non-standard and misspecified settings that make the doubly-nonparametric approach invaluable as it removes the need to correctly specify the response distribution from the outset.

Furthermore, for generalized additive models it is not at all easy to identify or postulate appropriate working distributions {\it a priori}. Marginal plots of the response against each covariate are incapable of showing the joint effect of the smooth additive predictors on the conditional mean and variance of the response. It is again in such scenarios that the doubly non-parametric approach proves invaluable, as a correct specification of the conditional variance is no longer needed.


For each set of simulations we consider sample sizes of $n = 200$ and $500$ with $N = 1,000$ replications. For continuous data settings, we consider a set of three popular working variance models, namely, $V(\mu) = \sigma^2$ (constant variance), $V(\mu) = \mu$ (linear variance) and $V(\mu) = \mu^2$ (quadratic variance). For count data settings, we consider the set $V(\mu)=\sigma^2$, $V(\mu) = \mu$ and $V(\mu) = \mu + \phi \mu^2$ (negative-binomial variance). For binary data settings, we consider the set $V(\mu) = \sigma^2$, $V(\mu) = \mu$ and $V(\mu) = \mu(1-\mu)$ (Bernoulli variance). These all correspond to ``classical" GAM settings and can be fit to data using the \texttt{gam} function from the \texttt{mgcv} R package \citep{W2016}. We also model each simulated dataset using the proposed doubly-nonparametric (DNP) GAM. All methods used cubic truncated P-splines \citep{RWC2003} with 10 knots placed at the deciles of the covariates. 
MATLAB code for fitting DNP GAM can be obtained by emailing the authors.

The values of the smoothing parameters used in each of the classical GAM approaches were automatically chosen by generalized cross-validation in the $\texttt{gam}$ function in R. For the doubly-nonparametric approach, we used the default smoothing parameters obtained from fitting a preliminary working $\texttt{gam}$ model to the data. For continuous data, this preliminary model was the normal $\texttt{gam}$ model. For count data, this preliminary model was the Poisson $\texttt{gam}$. For binary data, this preliminary model used was the Bernoulli $\texttt{gam}$. We again note that an advantage of carrying out the smoothing on the mean scale, rather than the canonical scale, is that the smoothing parameters become rather robust to the working distributional model. Although these smoothing parameters might not be ``optimal" for the doubly-nonparametric approach, this simple plug-in method emulates how one might actually approach each type of problem in practice. Moreover, simply using the default smoothing parameters given by a preliminary $\texttt{gam}$ fit to each dataset, rather than fine-tuning our method using knowledge of the true model, makes this a more-than-honest comparison with existing methods, and can also demonstrate the robustness of the proposed approach. In practice, researchers can directly cross-validating the doubly-nonparametric approach to get the ``optimal" smoothing parameters for each problem at hand.


Table \ref{tab:results} displays the average 95\% pointwise coverage rates of each smooth function and the overall mean curve across all observations, for simulations with sample size $n=200$. The results for sample size $n=500$ can be found in the Online Supplement -- these essentially confirm that the proposed method indeed approaches nominal coverage rates for the mean curve as sample size increases.


\begin{table}
\footnotesize
\centering
\caption{Average coverage rates (\%) for pointwise 95\% confidence bands for each smooth function and overall mean function, using GAMs with specified variance functions and doubly-nonparametric (DNP) GAMs. $N = 1,000$ simulations in each setting. Sample size $n=200$ for each simulation.}
\begin{tabular}{cclllllcclllll}
       & variance    & \multicolumn{5}{c}{1. Gamma}            && variance    & \multicolumn{5}{c}{2. Heteroscedastic Normal} \\
method & function   & $f_1$& $f_2$ & $f_3$ & $f_4$ & $\mu$ & & function   & $f_1$ & $f_2$ & $f_3$ & $f_4$ & $\mu$  \\
GAM     & $\sigma^2$     & 79.5 & 78.0  & 69.8  & 70.2  & 77.7  & & $\sigma^2$     & 87.0  & 85.5  & 79.9  & 76.0  & 84.4   \\
       & $\mu$      & 68.7 & 67.8  & 65.9  & 65.2  & 67.8  & & $\mu$      & \multicolumn{5}{c}{Not Applicable} \\
       &$\phi \mu^2$& 92.2 & 92.6  & 68.1  & 93.0  & 86.9  & & $\phi \mu^2$ &  \multicolumn{5}{c}{Not Applicable} \\
DNP    & ---        & 96.5 & 95.6  & 82.8  & 96.4  & 93.3  & & ---        & 86.3  & 86.0  & 83.0  & 83.5  & 85.3  \\
\\
       & variance    & \multicolumn{5}{c}{3. Poisson}              && variance    & \multicolumn{5}{c}{4. Negative-Binomial} \\
method & function   & $f_1$& $f_2$ & $f_3$ & $f_4$ & $\mu$ && function  & $f_1$ & $f_2$ & $f_3$ & $f_4$ & $\mu$ \\
GAM     & $\sigma^2$     & 85.8 & 85.2  & 74.0  & 76.7  & 82.9  & & $\sigma^2$     & 59.6  & 61.9  & 55.2  & 58.1  & 50.6   \\
       & $\mu$      & 94.3 & 94.5  & 78.7  & 93.5  & 90.6  & & $\mu$      & 82.6  & 81.2  & 73.8  & 80.1  & 80.2   \\
       & $\mu+\phi\mu^2$  &  95.0  & 95.2 &  75.3  & 94.1  & 90.6 & &$\mu+\phi \mu^2$ & 94.4 & 93.9  & 84.2  & 93.8  & 92.3   \\
DNP    & ---        & 91.8 & 91.6  & 80.2  & 89.6  & 89.0  & & ---        & 94.1  & 93.5  & 89.8  & 93.4  & 92.6  \\
\\
       & variance    & \multicolumn{5}{c}{5. COMPoisson (under-dispersed)}&& variance    & \multicolumn{5}{c}{6. COMPoisson (over-dispersed)}\\
method & function   & $f_1$& $f_2$ & $f_3$ & $f_4$ & $\mu$ & & function   & $f_1$ & $f_2$ & $f_3$ & $f_4$ & $\mu$ \\
GAM     & $\sigma^2$     & 91.2 & 90.5  & 76.8  & 85.5  & 87.5  & & $\sigma^2$     & 69.2  & 71.2  & 61.4  & 65.2  & 68.1   \\
       & $\mu$      & 99.1 & 99.2  & 82.3  & 99.0  & 97.0  & & $\mu$      & 86.5  & 86.2  & 75.3  & 85.3  & 84.0   \\
       & $\mu+\phi \mu^2$ & \multicolumn{5}{c}{Not Applicable} && $\mu+\phi \mu^2$ &  94.8 & 94.4 & 82.9 & 94.6 & 92.5 \\
DNP    & ---        & 91.3 & 91.5  & 76.0  & 89.5  & 88.0  & & ---        & 93.4  & 92.7  & 87.1  & 92.6  & 91.7   \\
\\
       & variance    & \multicolumn{5}{c}{7. Binomial}&& variance   & \multicolumn{5}{c}{8. Quasi-Binomial} \\
method & function   & $f_1$& $f_2$ & $f_3$ & $f_4$ & $\mu$ & & function   & $f_1$ & $f_2$ & $f_3$ & $f_4$ & $\mu$ \\
GAM     & $\sigma^2$     & 90.0 & 90.5  & 83.8  & 89.9  & 88.3  & & $\sigma^2$     & 89.8  & 90.5  & 83.8  & 89.8  & 88.5  \\
       & $\mu$      & 28.7 & 20.5  & 32.4  & 99.8  & 89.1  & & $\mu$      & 20.7  & 14.3  & 30.4  & 96.5  & 79.8  \\  
       &$\mu(1-\mu)$& 93.5 & 94.0  & 85.7  & 93.9  & 91.3  & &$\mu(1-\mu)$& 78.3  & 79.0  & 77.8  & 80.0  & 79.1  \\
DNP    & ---        & 92.4 & 92.4  & 86.6  & 92.8  & 90.5  & & ---        & 92.6  & 92.8  & 83.1  & 92.8  & 89.6   \\
\end{tabular}
\label{tab:results}
\end{table}

We see from Table \ref{tab:results} that the doubly-nonparametric approach can perform as well as correctly-specified models, even with suboptimal smoothing parameters. For misspecified models, its performance can be much better than classical approaches with incorrectly-specified response working variance functions. In particular, coverage rates for classical GAMs can be quite biased under model misspecification. The increased accuracy in inferences is due to the model flexibility induced by treating the error distribution as an infinite-dimensional parameter in the doubly-nonparametric framework. This reinforces the versatility and flexibility of exponential families for modelling data, as argued for in \citet{Hiejima1997}.

\begin{figure}[]
\includegraphics[width = 0.9\textwidth, trim = {2mm,  2mm , 2mm, 2cm}, clip]{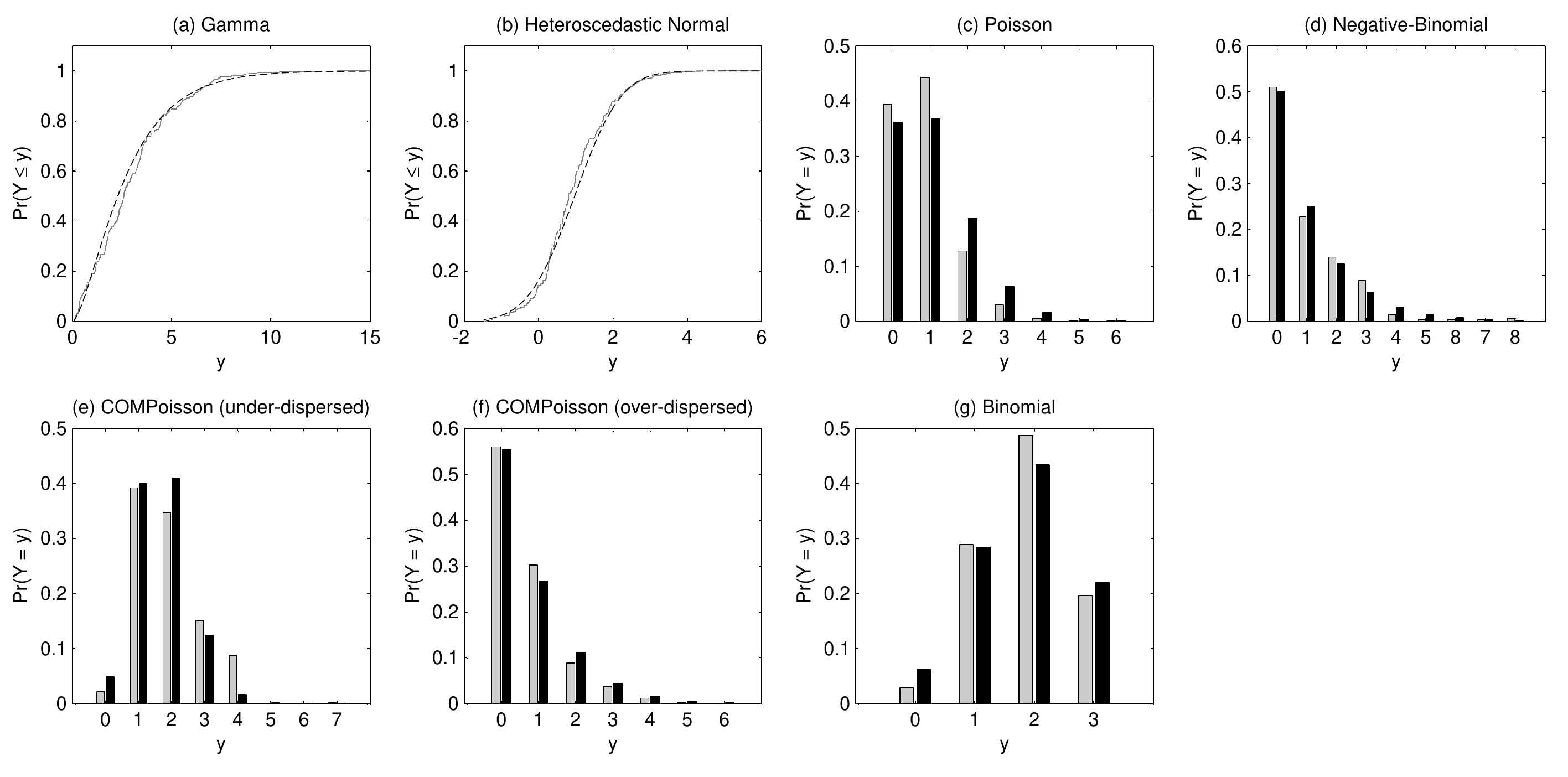}
\centering
\caption{Estimated (grey, solid) and true (black, dashed) distributions for each simulation setting. Sample size $n=200$.}
\label{fig:Fhat}
\end{figure}

As mentioned in Section \ref{sec:asy}, one of the key advantages of the doubly-nonparametric approach is that it also provides a consistent estimate of the underlying data-generating distribution. This estimated distribution can be plotted against parametric distributions for model selection and diagnostics in the classical GAM framework. To illustrate this, the estimated distributions $\hat F$ (or probability mass functions $d \hat F$ for discrete distributions) are plotted below against the true data-generating distribution from each of our parametric simulation settings. In each case, the estimated distribution closely matches the true underlying distribution.


\section{Data analysis example}
\label{sec:real}

\subsection{Divorce data}
We apply the proposed approach to model divorce rates in the US between 1920 and 1996 as a function of unemployment rate, female participation rate in labour force, births rate, military personnel rate and marriages rate. The rates are all measured in terms of number of cases per 1000 females. The dataset consists of 77 samples and comes from \texttt{faraway} R package \citep{Faraway2016}.

As divorce rates vary between 6 and 23 cases per thousand, a gamma GAM coupled with the log-link would be a reasonable model from a classical GAM point of view. For comparisons, we also fit the data using the DNP approach with a log link. Both approaches used 10-knot quadratic truncated P-splines to approximate each smooth functional predictor. 

\begin{figure}
\includegraphics[trim = 3.1cm 3.5cm 3.1cm 3cm, clip, scale = 0.6]{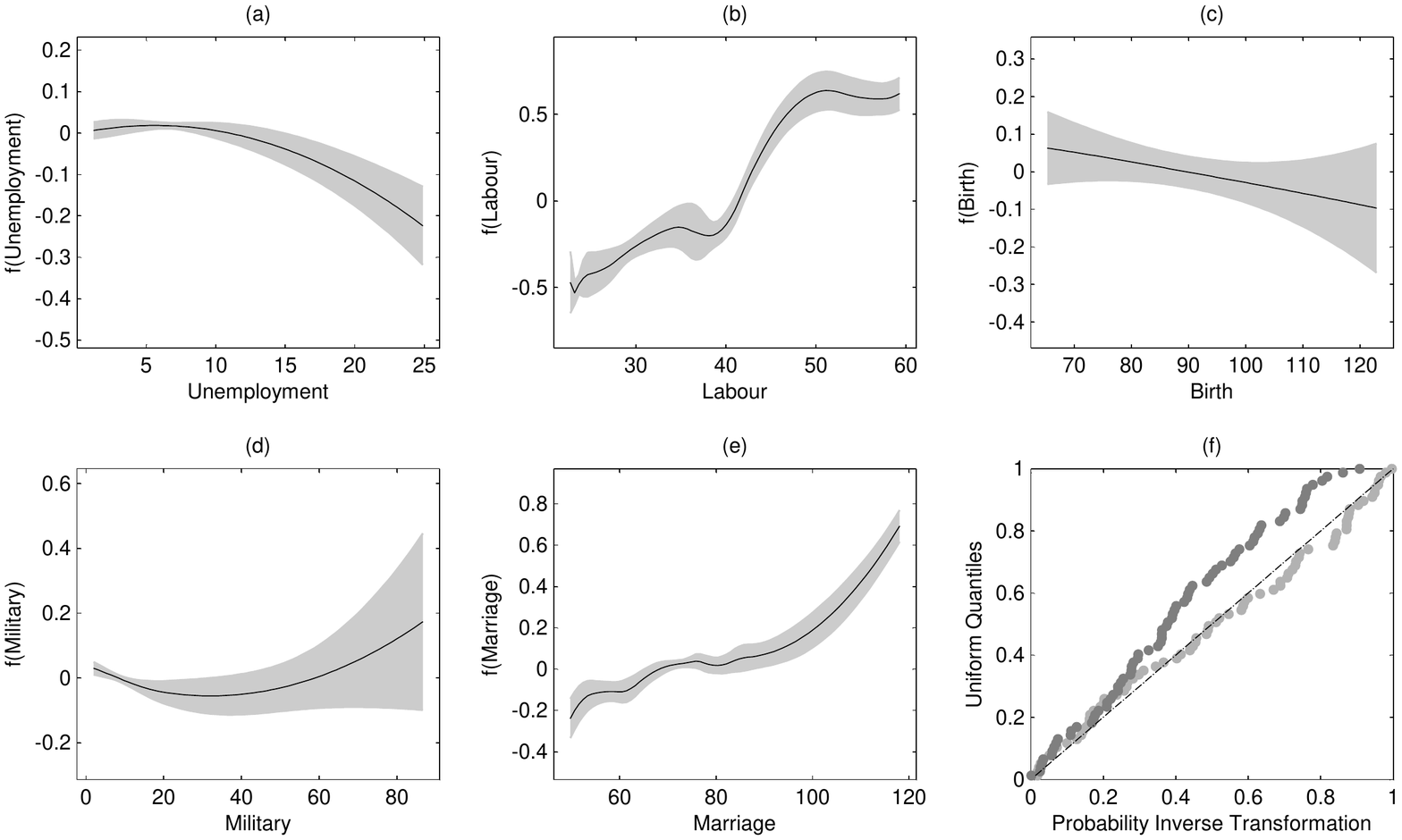}
\centering
\caption{Divorce rates data: (a)-(e) fitted curves (solid) and 95\% confidence bands (light grey, shaded); (f) PIT-uniform quantile plots for fitted DNP (light grey) and gamma (dark grey) GAMs.}
\label{fig:div}
\end{figure}

Figure \ref{fig:div} (a)-(e) displays the estimates of each curve along with their corresponding confidence bands using the DNP approach. We find that unemployment rates and birth rates have an overall negative association with divorce rates, while military, labour and marriage rates are generally positively associated with divorce rates. Also plotted in Figure \ref{fig:div} (f) is the PIT plot for the fitted model (light grey) as well as the PIT plot for the corresponding gamma GAM (dark grey).  We see that the doubly-nonparametric approach is a much better fit to the data than the gamma model, with the PIT of the former being very close to the uniform distribution. Moreover, the Kolmogorov-Smirnov test statistic for testing the gamma distribution is 0.694 with a $p$-value less than 0.001, confirming that the gamma model is indeed not a good fit for the data. Thus, model estimates and inferences based on the gamma distribution may well be biased, with the proposed doubly-nonparametric approach being a better fit for these data.  

\subsection{Science scores data}
We apply the proposed framework to model science scores as a function of income index, education index and health index across 52 countries. The science scores were obtained from the Programme for International Student Assessment, where students were assessed in science, mathematics, reading, collaborative problem solving and financial literacy\footnote{http://www.oecd.org/pisa}. The income index is measured by gross national income per capita, the education index is determined by the mean of years of schooling for adults aged 25 years and more and expected years of schooling for children of school entering age, and the health index is assessed by life expectancy at birth.\footnote{http://hdr.undp.org/en/content/human-development-index-hdi}

Residual plots from a preliminary Gaussian additive model analysis of the data\footnote{https://m-clark.github.io/docs/GAM.html} indicate fairly strong heteroscedasticity in the data. This renders the fitted model invalid, leading potentially to biased inferences on model components. In fact, it is rather difficult to determine an appropriate conditional variance function in this scenario, as it is not clear how the predictors jointly affect the conditional variability of the data. This is precisely when the proposed DNP approach proves invaluable, as it can relax such distributional assumptions and offers a certain flexibility and robustness to the underlying data-generating mechanism. 

To this end, we modelled the data with a DNP GAM using the identity link and 10-knot quadratic truncated P-splines. Figure 3 (a)-(c) displays the fitted curves for each covariate, along with their corresponding confidence bands. We see that while education and health have an overall monotonic relationship with science scores, the effect of income does not appear to be monotonic. Indeed, the relationship between the scientific performance of students from different countries and their national wealth seems to be rather complex. If we fix both the education and health index, it is not surprising that students get considerably lower scores from impoverished countries as these countries may do not have sufficient money for the national education. However, wealthy countries in terms of their gross national income also do not guarantee higher science scores. The proportion of investment in education and many other neglected factors may need to be introduced into the model to provide a better understanding of these results.

Finally, the PIT plot in Figure 3 (d) confirms that the fitted model is indeed appropriate for these data. The DNP approach has adequately accounted for the heteroscedasticity in the data in a completely nonparametric way.

\begin{figure}
\includegraphics[trim = 0cm 7cm 1cm 6cm, clip, scale = 0.6]{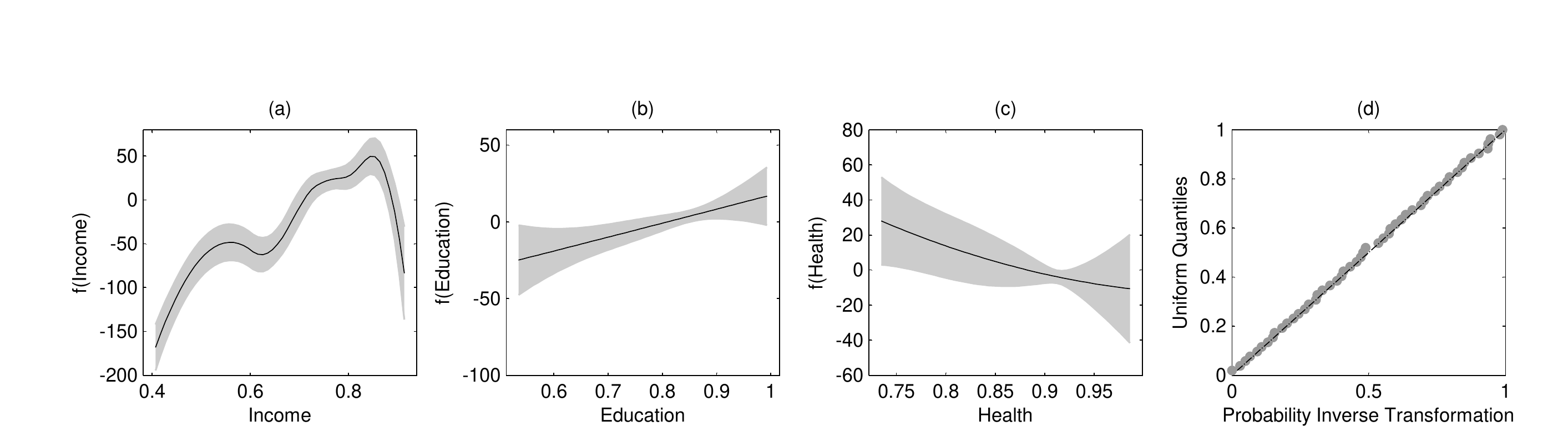}
\centering{\caption{Science scores data: (a)-(c) fitted curves (solid) and 95\% confidence bands (shaded); (d) PIT-uniform quantile plots for fitted DNP GAM.}}
\label{fig:sci}
\end{figure}

\section{Discussion}
\label{sec:summary}
The confidence bands in this paper are constructed from a frequentist approach. The finite-sample performance of the proposed method, although a marked improvement over misspecified models, may still be biased due to the penalty-induced bias problem as noted in \citet{W2006}. This may be improved by considering a corresponding Bayesian approach, similar to that in \citet{MW2012}. In addition, constructing simultaneous confidence bands for each unknown function and for the overall mean curve would be topics for future research.

\appendix
\section*{Appendix}
The results in Section \ref{sec:asy} hold under the regularity conditions below.

\begin{enumerate}[label=A\arabic*.]\setlength\itemsep{-0.1em}
  \item The response space $\mathcal{Y}$ is (contained in) a closed, finite interval $[L, U]$ in $\mathbb{R}$ and the covariate space  $\mathcal{X}$ is (contained in) a closed, finite hyperrectangle in $\mathbb{R}^d$.
  \item There exists $\delta_1 > 0$ such that $\mu(\bm{x})$ maps into $\mathcal{Y}$ and $\mu'(\bm{X})$ and $\mu''(\bm{X})$ exist and are continuous on $\mathcal{X} \times \{\bm{\beta} \in \mathbb{R}^{K d} : ||\bm{\beta} - \bm{\beta}_*|| \leq \delta_1\}$.
  \item There exists $\delta_2 > 0$ such that $V(X; \bm{\beta}, F) \geq V_2$ on $\mathcal{X} \times \{(\bm{\beta},F) \in \mathbb{R}^{K d} \times \mathcal{F}_\mu: ||\bm{\beta} - \bm{\beta}_*, F - F_* || \leq \delta_2 \}$.
\end{enumerate}

\section*{Acknowledgements}

\section*{Supplementary material}
The Online supplement contains further simulation results and the technical details of Proposition 1.

\section*{References}

\end{document}